# Surface-Enhanced Raman Detection of the $CO_2$ Moisture Swing

Javier Mendez Lozoya, Estrella Solis Mata, J. Jesus Velazquez Salazar, Alondra Hernandez Cedillo, Miguel Jose Yacaman, and Jennifer L. Wade*

ABSTRACT: The development of scalable, energy-efficient carbon dioxide ($CO_2$) capture technologies is critical for achieving net-zero emissions. Moisture swing (MS) sorbents offer a promising alternative to traditional thermal regeneration methods by enabling reversible $CO_2$ binding through humidity-driven ion hydrolysis. In this study, we investigate the anion speciation dynamics in two classes of MS materials—an anion-exchange resin with a bicarbonate anion and activated carbon impregnated with potassium bicarbonate salt—using both sorption measurements and in situ surface-enhanced Raman spectroscopy (SERS). Ni-coated Ag nanowires were employed as SERS substrates to enhance signal intensity and enable the real-time detection of carbonate ($CO_3^{2-}$), bicarbonate ($HCO_3^-$), and hydroxide ($OH^-$) species under controlled humidity conditions in both air and nitrogen atmospheres. The results reveal humidity-dependent interconversion between anionic species with significant spectral shifts confirming the reversible hydrolysis reactions that drive the MS mechanism. Under humid conditions, we observed the depletion of bicarbonate signals and a concurrent increase in carbonate species, consistent with moisture-induced desorption of $CO_2$. With the activated carbon samples, we further observed the formation of hydroxide. These findings not only validate the mechanistic models of humidity-driven anion exchange in MS sorbents but also demonstrate the practical potential of SERS as an operando diagnostic tool for monitoring $CO_2$ capture media. The ability to resolve and semiquantitatively evaluate the reversible transformation of carbonate, bicarbonate, and hydroxide ions under realistic environmental conditions provides valuable insight for the rational design, performance optimization, and quality control of next-generation sorbent materials for direct air capture applications.

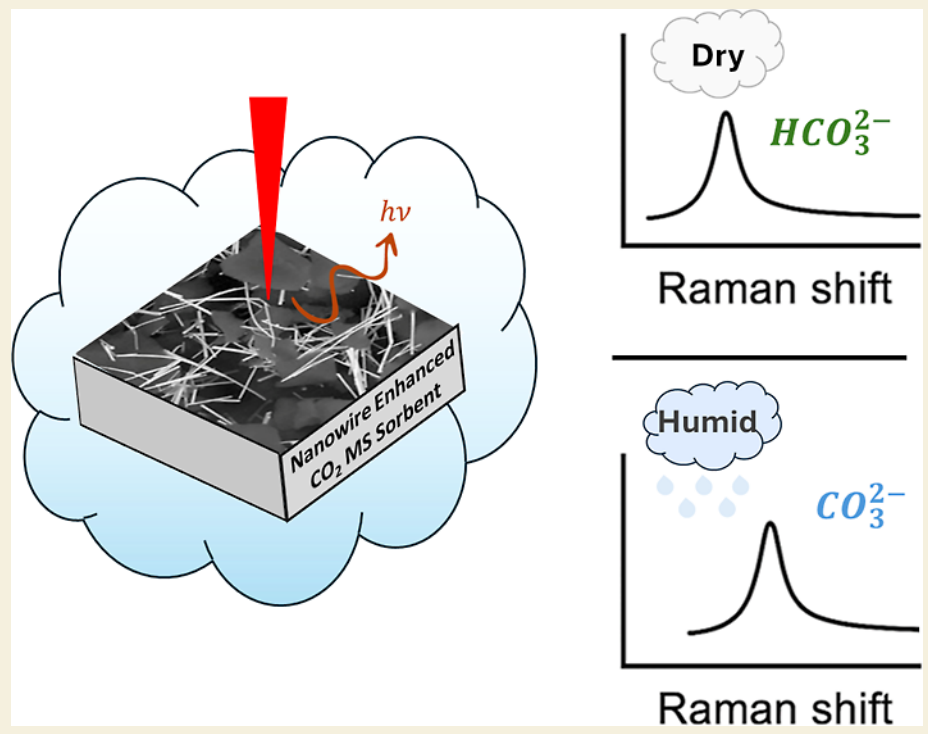

## 1. INTRODUCTION

Scalable and affordable carbon dioxide ($CO_2$) capture from dilute gas streams has grown in emphasis over the past decade to achieve both negative greenhouse gas emissions[1,2] and a circular carbon economy.[3] Emerging commercial technology for the separation of $CO_2$ from air, termed direct air capture (DAC), is based on hydroxide solvents[4] or supported amine sorbents.[5] The energetics of both approaches are dominated by the thermal energy required to regenerate the phase separating media and in turn concentrate the $CO_2$ for subsequent compression and storage or use.[6] The solid sorbent system has potential to drive down energetic costs through waste heat integration or renewable power heat pumps, as regeneration temperatures for these materials range from 80 to 120 °C versus the hydroxide system requiring >900 °C regeneration of the trapped $CO_2$ species, solid $CaCO_3$. However, the sorbent's $CO_2$ separation capacity is hampered by the thermal oxidative stability of the amine chemistry, limiting the amount of $CO_2$ captured per unit of amine sorbent.[7−9] Beyond the early commercial approaches, other promising technologies include the use of metal oxide sorbents, also requiring high-temperature regeneration (>900 °C),[10] and amino acid solvents,[11] also prone to thermal degradation.[12]

The work in this study examines an emerging family of $CO_2$ separating materials that use hydration rather than heat to regenerate and concentrate $CO_2$, the moisture swing (MS).[13−22] All MS-responsive materials provide nanoconfined domains that constrain reactive anions like anion exchange resins,[13,19,20,22,23] anion exchange membranes,[16] quaternized graphene sheets,[17] or with simple salt impregnated into microporous structures.[15,18,21] Guided by DFT studies[23−25] and MS measurements of $CO_2$ sorption across a range of alkaline anions, the central hypothesis behind the MS separation is that anion hydrolysis varies with humidity. The $CO_2$-based MS relies on the hydrolysis reactivity of the alkaline carbonate, $CO_3^{2-}$anion as a function of hydration, which is in turn controlled by the surrounding water activity i.e., relative humidity (RH) (eq 1).

$$CO_3^{2-}\cdot m_1H_2O \leftrightarrow HCO_3^{-}\cdot m_2H_2O + OH^{-}\cdot m_3H_2O + (m_1 - m_2 - m_3 - 1)H_2O \quad (1)$$

In the above expression, the divalent carbonate can dissociate water from its hydration cloud ($m_1H_2O$) to form the conjugate monovalent bicarbonate ($HCO_3^-\cdot m_2H_2O$) and hydroxide species ($OH^-\cdot m_3H_2O$). Notably, this hydrolysis becomes energetically favored at low hydration levels, where the monovalent anions are stabilized relative to the divalent carbonate. The amount of water released is dependent on the stoichiometric size of the hydration clouds ($m_iH_2O$), which is dictated by the affinity of water to the anion−cation pair and the affinity of water to the surrounding structure at a given level of water activity. In the presence of $PCO_2$, the hydroxide will readily bind this acid gas to form a further bicarbonate anion (eq 2).

$$CO_2 + OH^-\cdot m_3H_2O \leftrightarrow HCO_3^-\cdot m_2H_2O \quad (2)$$

Taking eqs 1 and 2 together, the net chemical reaction of $CO_2$ binding can be represented as eq 3, where the hydroxide species is treated as a transitory player.

$$CO_3^{2-}\cdot m_1H_2O + CO_2 \leftrightarrow 2HCO_3^-\cdot m_2H_2O + (m_1 - m_2 - m_3 - 1)H_2O \quad (3)$$

The overall result is a net release of water upon binding of $CO_2$, where $m_1$-$m_2$-$m_3$-1 > 0. Consequently, introducing water to the system shifts eq 3 to the left, liberating $CO_2$ and resulting in the MS in accordance with Le Chatlier's principle. Under low $PCO_2$, the equilibrium in eq 2 favors the hydroxide form. This latter point has been shown both through MS-based biocomponent isotherms[26,27] and dynamic breakthrough column experiments,[23] where total $CO_2$ (de)sorption matches the ion exchange capacity of the anion exchange resin when desorption proceeds under low, $PCO_2 < 1$ Pa, and in alkaline electrolyte studies where hydroxide-exchanged anion exchange films remain in the OH- state only under an inert atmosphere (e.g., in UHP $N_2$).[28,29] Thus, it is the influence of both water activity (hydration) and $PCO_2$ that dictates the anion speciation in these confined materials.

Most MS investigations have evaluated $CO_2$ and $H_2O$ sorption through gas phase changes in water vapor and $CO_2$. Fewer studies have experimentally interrogated the reactive species to validate the above mechanism. Two exceptions have been the use of solid-state NMR to probe the carbonate/bicarbonate species after equilibration at different levels of moisture.[17,30] Nicotera et al. examined the MS mechanism in trimethylammonium-functionalized graphene oxide using $^{13}C$ NMR, and there they saw the depletion of $HCO_3^-$ and formation of $CO_3^{2-}$ when equilibrating to 400 ppm[13] $CO_2$ at various humidity levels.[17] This study did not explore the formation of $OH^-$, which was the initial anion in their material. Yang, Singh, and Schaefer 2018 similarly used $^{13}C$ NMR methods to monitor anion speciation from a polystyrene, divinylbenzene cross-linked anion exchange resin,[30] also initialized in the hydroxide form. Yang et al. also observed the formation of $HCO_3^-$ after binding with $CO_2$; however, the investigators did not see the emergence of a $CO_3^{2-}$ phase upon the addition of humidity. Rather they observed the formation of hydroxide, suggesting a different reactive pathway than the one proposed by eq 3. Details of the $PCO_2$ were not reported in this latter case. A clear understanding of the MS reactive mechanism is essential in informing future material advancements that aim to increase $CO_2$ capacity, sorption kinetics, or material stability.

The goal of this work is to examine the anion speciation in two classes of MS-responsive materials: a commercial anion exchange resin, IRA900, and an activated carbon impregnated with potassium carbonate salts using in situ Raman spectroscopy. The use of an environmental chamber interfacing with a Raman laser and detector enables a step change in water activity surrounding the sorbent materials while in the presence (~400 ppm $CO_2$ in air) or absence of $CO_2$ (in ultrahigh purity $N_2$). This in turn enables the detection of emerging or depleting anion phases as they react with the surrounding atmosphere.

Raman spectroscopy, based on inelastic light scattering, enables qualitative identification of molecular species and quantitative determination of analyte concentrations, making it ideal for evaluating the reactive anion of anion-exchange resins.[31−39] Surface-Enhanced Raman Spectroscopy (SERS) leverages localized surface plasmon resonance and surface plasmon polaritons on nanostructured Ag, Au, or Cu substrates to enhance electromagnetic fields by several orders of magnitude, enabling trace-level detection and real-time monitoring of adsorption−desorption dynamics critical to assessing resin performance.[40−48]

In the present study, we utilize Ni-coated Ag nanowires (NWs) as SERS-active substrates. These nanowires, previously characterized in earlier work,[49] combine the strong plasmonic activity of silver with the magnetic functionality of nickel. The one-dimensional anisotropic geometry of the nanowires facilitates the generation of localized electromagnetic "hot spots," especially at junctions and tips, leading to significantly enhanced Raman signal intensities. The Ni shell provides magnetic responsiveness, enabling facile manipulation and potential substrate reusability and thus enhancing experimental reproducibility. Additionally, the large surface area and high structural integrity of Ni coated Ag NWs promote effective analyte adsorption and long-term stability, rendering them excellent candidates for sensitive and reproducible SERS-based detection.

## 2. METHODS

### 2.1. Materials

The MS materials explored in this study include a divinylbenzene cross-linked polystyrene anion exchange resin with trimethylammonium groups charge-balanced in the chloride form (IRA900-Cl), and potassium bicarbonate-impregnated activated charcoal (AC-$KHCO_3$). All raw sorbent materials were purchased from Sigma-Aldrich: IRA-900-Cl (Amberlite), activated charcoal (Spectrum Chemicals), and potassium bicarbonate salts (ACS reagent, >99%). The IRA900 resin is received as a 0.5−1 mm spherical bead. To enhance contact between the resin and a nanowire, the resin was cryomilled using a SPEX SamplePrep 6870 cryogenic milling machine, with a milling cycle of 5 min, followed by a 2 min cooling period in liquid nitrogen to dissipate heat buildup. This cycle was repeated six times. To activate the IRA900-Cl to a MS responsive state, the chloride must be ion-exchanged to a reactive form. For this study, the IRA900-Cl was ion-exchanged using two 24 h washes in 0.1 M $KHCO_3$ solution at a ratio of 15 mL liquid:20 mg resin. Excess potassium salts were rinsed from the resin using deionized (DI) water. The concentration of exchanged $Cl^-$ to $HCO_3^-$ was measured by using Hach chloride strips in the ion-exchange solution. After two washes the net ion exchange capacity was $3.0 \pm 0.1$ mmol/g resin, where the numerator represents the dry mass of the IRA900-Cl after sitting in room air for 24 h (∼15% RH, 20 °C). Both the bicarbonate-exchanged resin (*IRA*900-$HCO_3$) and the as-received chloride resin (IRA900-Cl) were evaluated in this study, with the latter acting as a control to compare results to the reactive form. The porous activated charcoal samples were wet impregnated with 1.0 M $KHCO_3$ following the procedure detailed by Zheng et al.[18] The Ni-coated Ag nanowires used for the surface enhancement were prepared using methods described elsewhere.[49]

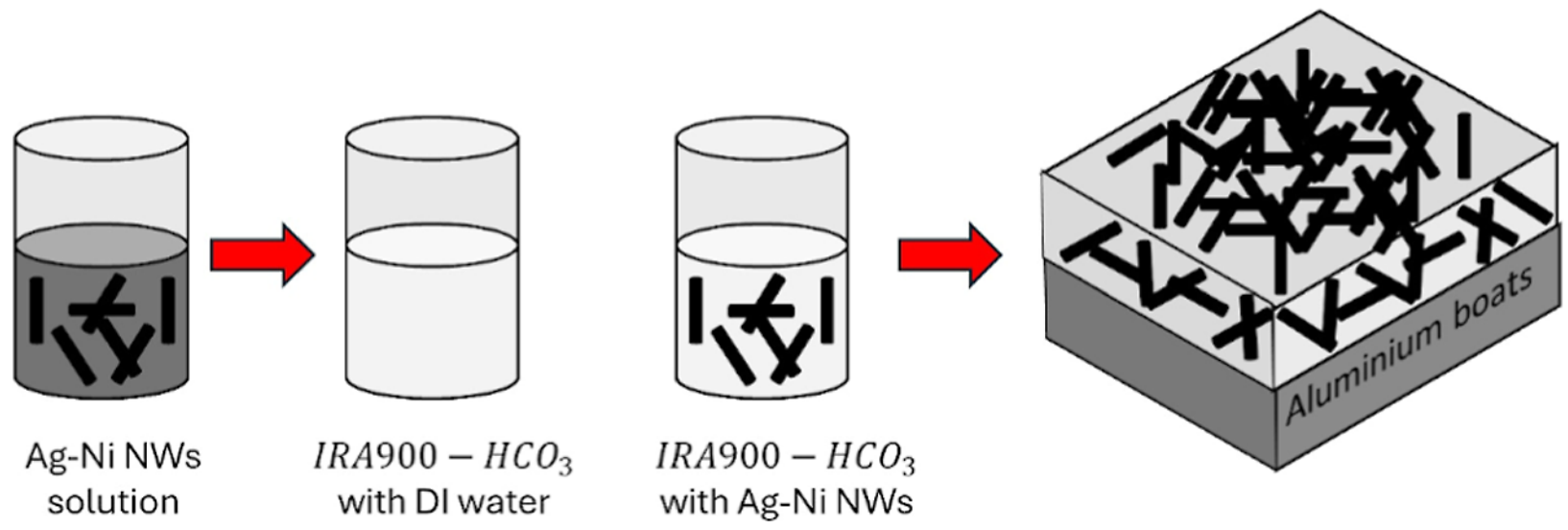

**Figure 1.** Diagram illustrating the fabrication process of SERS substrates.

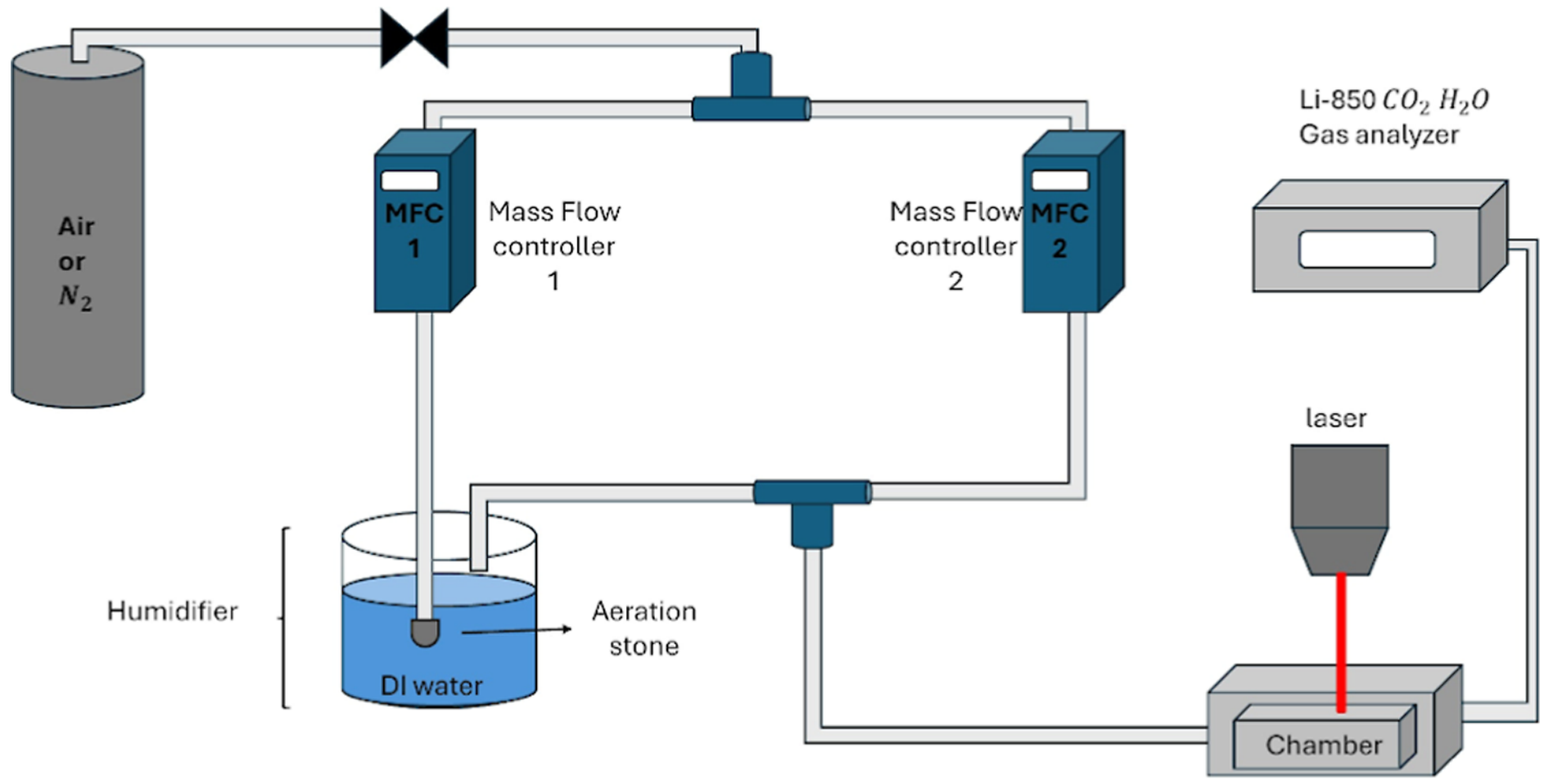

**Figure 2.** Diagram of the Raman measurement process under dry and wet conditions.

## 2.2. Surface-Enhanced Sample Preparation

We impregnated 50 $\mu$L of a cryomilled IRA900-Cl or *IRA*900-$HCO_3$ solution with 50 $\mu$L of a silver−nickel core−shell nanowire (Ni-coated Ag NW) suspension. To ensure a homogeneous mixture suitable for deposition and analysis, we dispersed both components in deionized water. In a similar approach, we impregnated salt-treated *i*-$KHCO_3$ samples with 50 $\mu$L of a Ni-coated Ag NWs suspension. We used a 1 M $KHCO_3$ solution as the dispersion medium to retain the reactive salt species on the carbon surface during impregnation. The resulting suspensions were homogenized using a vortex mixer, deposited onto aluminum boats, and air-dried overnight prior to further analysis (Figure 1).[36]

## 2.3. Material Characterization

The morphology of the Ni-coated Ag nanowires (NWs) and IRA900-$HCO_3$ resin was characterized using a Phenom ProX scanning electron microscope (SEM) equipped with a backscattered electron (BS) detector, operated at an accelerating voltage of 10 kV, and a Talos F200i scanning transmission electron microscope equipped with a bright field detector and operated at 200 kV. Elemental composition of the nanowires was examined via energy-dispersive X-ray spectroscopy (EDS) mapping.

Further, the moisture-swing-based $CO_2$ and $H_2O$ countersorption of the sorbents (void of nanowire) was characterized by comparing sorption behavior relative to the speciation observed in Raman measurements. We monitored this mixed gas sorption using a modified Netzsch STA 449 F3 Jupiter instrument, which combines thermogravimetric analysis (TGA), differential scanning calorimetry, and outlet gas analysis (Li-7000 infrared $H_2O/CO_2$ analyzer) with control of the gas composition in an open flow configuration. This contrasts with most MS characterizations explored in a closed flow configuration where only changes in gas composition are controlled and monitored.[15,19,21,27] Experiments were performed under 25 °C isothermal conditions, and step-changes in humidity were controlled using a modular humidity generator system (MHG32, Prohumid). Sorbents were equilibrated for 5 h at 25 °C in a 20% RH humidified stream of 400 ppm of $CO_2$ in ultrahigh-purity nitrogen ($N_2$) with a total dry gas flow of 200 sccm. At 5 h intervals, the water activity was stepped between 95 and 20% RH, and the corresponding changes in sorbent mass, heat flow, and gas composition were recorded. Total sorbed $CO_2$ over time was calculated from the integration of the outlet $CO_2$ gas concentration signal relative to the final 60 min baseline of the 5 h segment. In contrast, total sorbed $H_2O$ over time was calculated from the net mass change measured with the TGA microbalance, corrected for the mass of $CO_2$ (de)sorbed in each step. Further details on the instrumentation, gas flow configuration, and mass balance calculations are described elsewhere.[26]

## 2.4. In Situ Raman Environmental Chamber

Moving beyond gas and mass sorption analyses, we directly probed the compositional changes of the sorbent materials under varying environmental conditions using surface-enhanced Raman scattering (SERS) spectroscopy. In situ Raman measurements were taken over time in open-flow, dry, and humidified conditions using either compressed air, which contained approximately 450 ppm $CO_2$, or ultrahigh-purity nitrogen ($N_2$). Prior to introducing water vapor, the dry gas passed through the line connected to second mass flow controller at a rate of 100 sccm for 3 h, drying the sorbent sample. The outgoing feed gas was then directed to a Li-850 $CO_2/H_2O$gas analyzer, where the concentrations of $H_2O$ (in PPT) and $CO_2$ (in PPM) were measured. After drying, the gas was switched to flow through MFC1, which directed the flow into a humidifier, where the gas was sparged into DI water using an aeration stone. Figure 2 shows the schematic configuration used to make the Raman measurements using Horiba XploRA PLUS equipment. Raman spectra were acquired using a red laser with a wavelength of $\lambda$ = 638 nm, an 1800 $cm^{-1}$ grating, and a 10× Nikon objective lens. The incident laser power was set to 10 mW, and spectra were collected over the range 400−2000 $cm^{-1}$ for the polymer resins. This range was selected to minimize potential photothermal damage associated with the use of a red laser excitation source. Lower-wavenumber regions, typically below 400 $cm^{-1}$, are more susceptible to fluorescence interference and thermal degradation, particularly in carbon-based or salt-impregnated materials. By restriction of the measurement range, the structural integrity of the sample was preserved while capturing the most relevant vibrational modes for compositional and structural characterization. Raman spectra of the AC-$KHCO_3$ samples were acquired using a green laser excitation source with a wavelength of $\lambda$ = 532 nm. The measurements were performed with a grating of 1800 $cm^{-1}$ and a 10× Nikon objective lens, providing adequate spatial resolution for surface analysis. A laser power of 10 mW was employed to ensure sufficient signal intensity while avoiding thermal degradation of the sample. Spectral data were collected over the range of 400−3800 $cm^{-1}$, enabling comprehensive analysis of both low- and high-frequency vibrational modes associated with the carbon framework and surface-bound species. For each sample, 27 Raman spectra were recorded under both dry and humid conditions in air and nitrogen gas atmospheres. Averaging across these replicates enabled the generation of representative spectra for each condition, minimizing the impact of instrumental noise and sample variability on spectral interpretation.

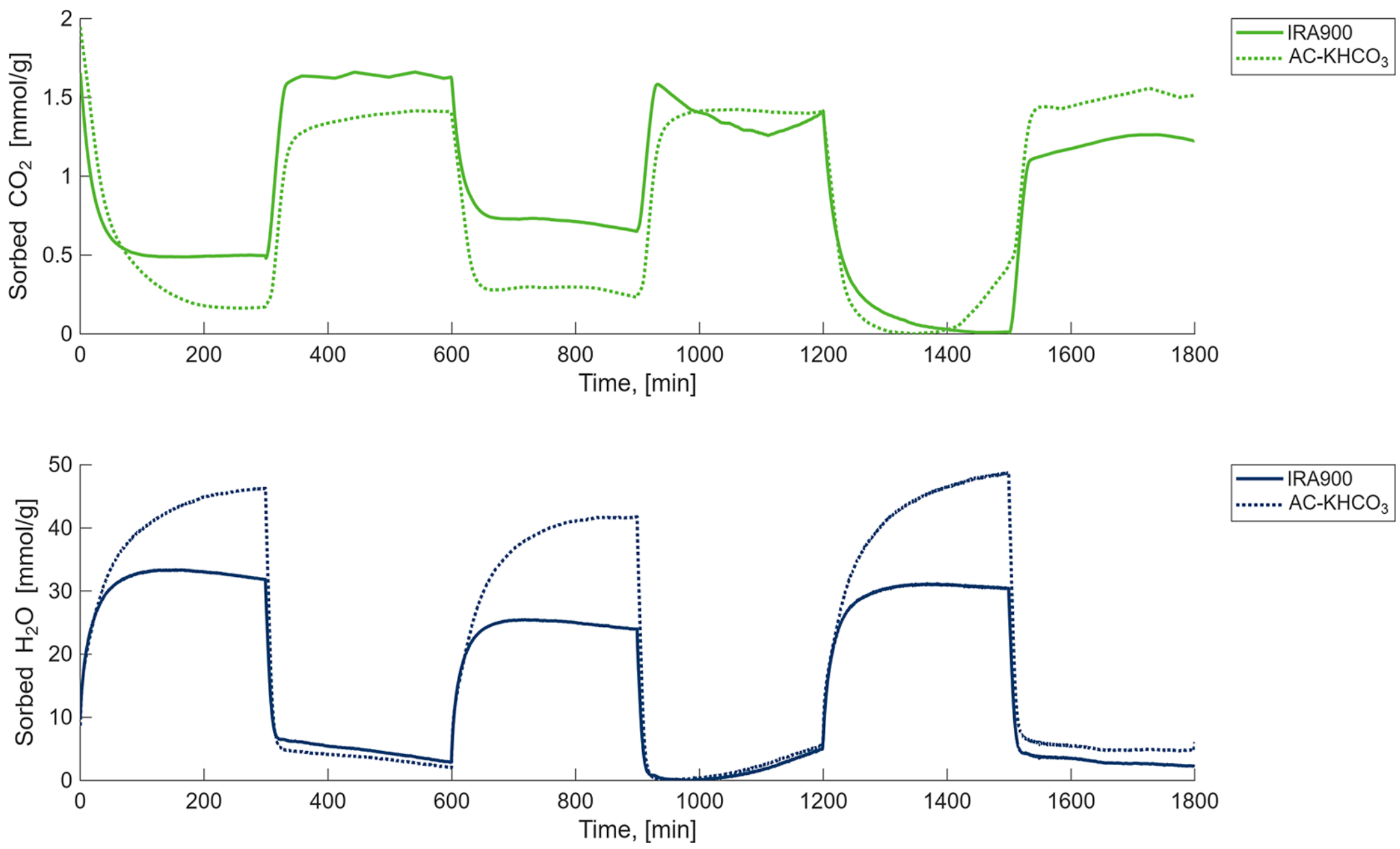

**Figure 3.** Sorption of $CO_2$ (green) and $H_2O$ (blue) on the IRA900-$HCO_3$ (solid) and AC-$KHCO_3$ (dashed) sorbents. Experiments were conducted at 25 °C with 40 Pa inlet $CO_2$ partial pressure in $N_2$. Step increases in water loading are associated with an increase in RH to 95%, decreases in water loading are associated with a decrease in RH to 20%.

### 2.5. Raman Spectral Analysis

Prior to spectral analysis, all Raman spectra were preprocessed to enhance signal quality and ensure reliable quantification of the anionic species. Initially, spectral smoothing was performed using the Savitzky−Golay method with a second-order polynomial and a 7-point window. This approach effectively reduced high-frequency noise while preserving the integrity of key spectral features such as narrow vibrational peaks associated with bicarbonate and carbonate species. Subsequently, baseline correction was applied to each spectrum to eliminate contributions from background fluorescence and instrumental drift, thereby improving the peak resolution and quantification accuracy. To account for variations in laser power, sample orientation, and acquisition conditions, the spectra were normalized to peaks corresponding to invariant features, serving as an internal standard for spectral normalization. For IRA900, the spectra were normalized to the peak intensity at 1613 $cm^{-1}$, corresponding to the C═C stretching mode of aromatic rings found within the polystyrenic backbone. For AC-$KHCO_3$, spectra were normalized to the peak intensity at 1350 $cm^{-1}$, corresponding to the D1-band.[50] Semi-quantification of bicarbonate ($HCO_3^-$) and carbonate ($CO_3^{2-}$) was performed by integrating the corresponding Raman bands centered at approximately 1017 $cm^{-1}$ and 1065−1068 $cm^{-1}$, respectively, and then normalized to the total spectral area. For the AC-based materials we also examined the hydroxide/water region of the Raman spectra to evaluate any hydroxide species either hydrogen bonded with water (2864 $cm^{-1}$) or free (3706 $cm^{-1}$).[43] To ensure consistency across all vibrational modes, the $OH^-$ band was treated as an individual component within the global fit and its peak area was extracted using the same fitting constraints and optimization parameters. The integrated area of the $OH^-$band was subsequently normalized using the same criteria as other bands, to either the total spectral area or the 1350 $cm^{-1}$ D1-band.

To ensure a good spatial representation of the sample, the Raman spectra were averaged over 27 independent scans acquired at different locations across the sample surface. This methodology captures spatial heterogeneity within the sample and provides a statistically robust representation of the bicarbonate-to-carbonate transformation under the experimental conditions.

## 3. RESULTS AND DISCUSSION

In the following section, we evaluate two families of moisture-responsive $CO_2$ sorbents using open-flow mixed gas sorption and in situ Raman spectroscopy in the presence of dilute $CO_2$ (~400 ppm) and its absence. Here we explore the behavior of a bicarbonate-exchanged macroporous anion exchange resin (IRA900-$HCO_3$) and an activated carbon-impregnated with potassium bicarbonate salts (AC-$KHCO_3$). The goal of this work is to interrogate the MS mechanism by identifying the dominating reactive species during changes in RH across two different families of materials that exhibit reversible MS.

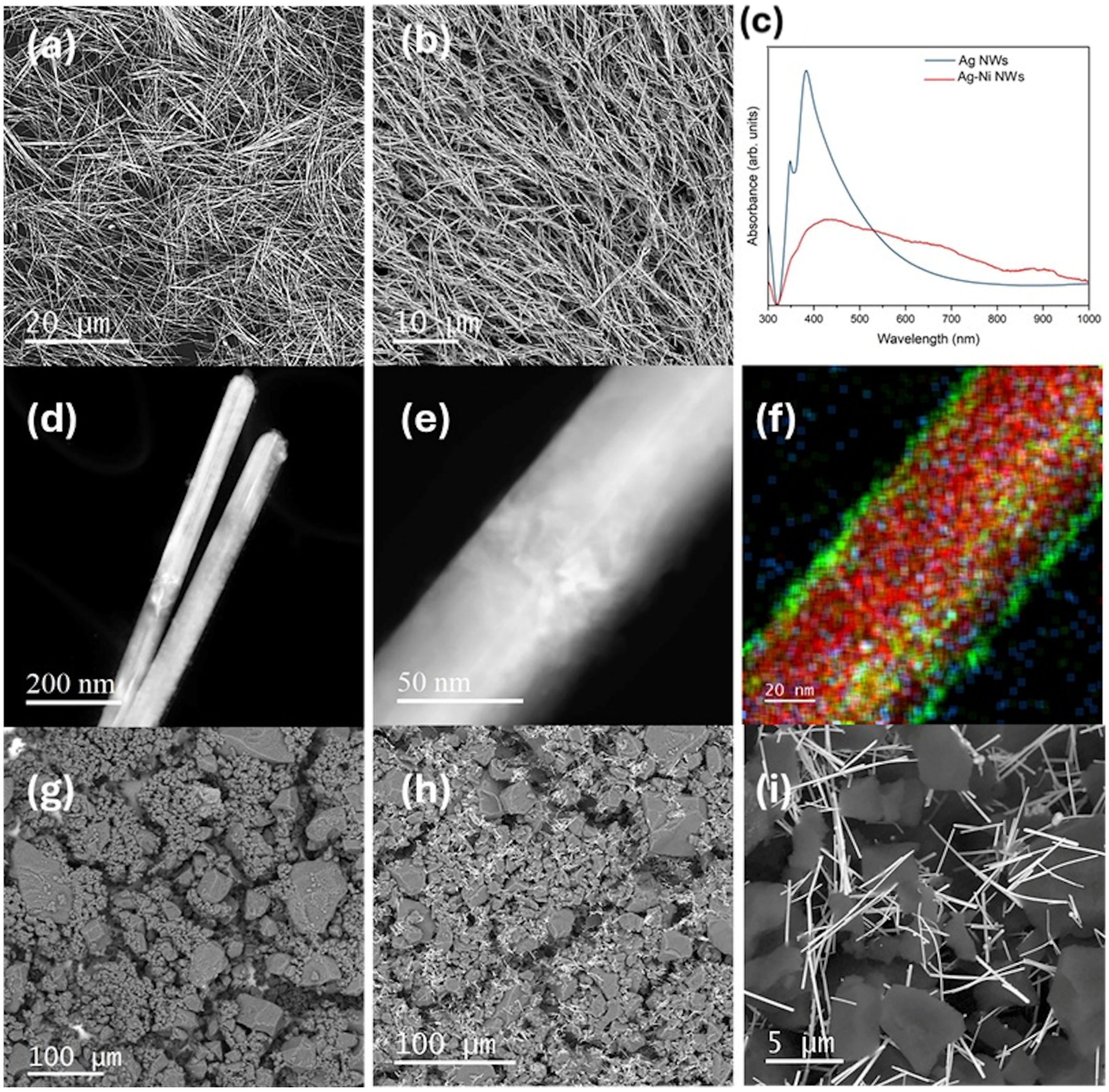

**Figure 4.** (a) SEM micrograph illustrating the distribution of Ag NWs. (b) SEM micrograph depicting the Ni-coated Ag NWs used in the fabrication of the SERS substrate. (c) UV−vis absorbance spectra of Ag nanowires (blue) and Ni-coated Ag nanowires (red). (d) Low-magnification TEM image showing uniform, elongated Ni-coated Ag nanowires with diameters of ∼80−120 nm. (e) High-resolution TEM (HRTEM) image of the nanowire. (f) EDS elemental mapping of a single nanowire confirming the spatial distribution of silver (red) confined to the core and nickel (green) forming a conformal shell. (g) SEM micrograph of the cryomilled *IRA*900-$HCO_3$ sample. (h) SEM micrograph of the *IRA*900-$HCO_3$ sample containing nanowires. (i) Higher magnification SEM micrograph of the *IRA*900-$HCO_3$ sample with wires, providing a detailed view of the distribution of Ni-coated Ag NWs across the sample.

### 3.1. $CO_2$ MS

Calculated cyclic sorbent loadings of $CO_2$ (green) and $H_2O$ (blue), driven by step changes in water activity that oscillate between 20 and 95% RH, are given in Figure 3. The inlet $CO_2$ partial pressure was fixed at 40 Pa, and the sorbent temperature was held isothermal at 25 °C. The outlet gas compositions, mass, and heat flows associated with these changes are provided in Figure S1 in the Supporting Information. Both sorbent materials, IRA900-$HCO_3$ (solid line) and AC-$KHCO_3$ (dotted line), resulted in an increase in water loading from the increase in humidity, as expected. However, the water loading on AC-$KHCO_3$ was greater than the polymer resin: 48 ± 2 vs 30 ± 4 mmol/g, where error represents the standard deviation of sorption over three cycles. Further, the rate of higher water loading in the AC-$KHCO_3$ did not reach an equilibrium level after 5 h. This higher loading is expected given the higher charge of the impregnated salt (e.g., 5.7 vs 3.0 mmol/g) and the finer micropores comprising activated carbon materials.[51,52]

As water bound to the sorbent materials, $CO_2$ was desorbed and released into the gas phase (see the $CO_2$ gas concentration alongside the sorbent mass change in Figure S1). Thus, $CO_2$ desorption is counter to $H_2O$ sorption in the MS process. The calculated $CO_2$ loading for both IRA900-$HCO_3$ and AC-$KHCO_3$ was similar to 1.1 ± 0.2 and 1.4 ± 0.4 mmol/g, respectively. Again, uncertainties represent the average working capacity of the $CO_2$ sorption across the three MS cycles. The ratio of $H_2O$ cycled per unit of $CO_2$ separated is thus greater for the IRA900 materials due to the smaller water loadings relative to $CO_2$ (11:1 wt/wt vs 13:1 wt/wt $H_2O$: $CO_2$). Note, the unsmooth plateau of the $CO_2$ loadings is an artifact of a drifting $CO_2$ gas concentration baseline during those respective cycles, and baseline drift of the $H_2O$ loading was due to mass balance drift during the experiment resulting in ∼5% error of the $H_2O$ loading (Figure S1).

As mentioned in Section 2.1, IRA900 was ion-exchanged to 3.0 mmol/g $HCO_3^-$. Upon hydration, bicarbonate is expected to convert to carbonate as $CO_2$ desorbs, resulting in a 1:2 ratio of $CO_2$ release per pair of charge sites or a theoretical loading of 1.5 mmol/g. At 1.1 mmol/g the sorbent cycles 73% of the theoretical capacity when oscillating between 20 and 95% RH. The bicarbonate impregnation into the AC was 5.7 mmol/g, which leads to a theoretical capacity of 2.9 mmol/g. Only 49% of the theoretical capacity was achieved in the impregnated solid. Thus, in $CO_2$-containing atmospheres, we expect to see a partial transition from bicarbonate to carbonate anions in these materials. In the absence of $CO_2$ (e.g., in UHP $N_2$), the bicarbonate species may also decompose to hydroxide, according to eq 2. All three possible reactive anion species are examined by using Raman.

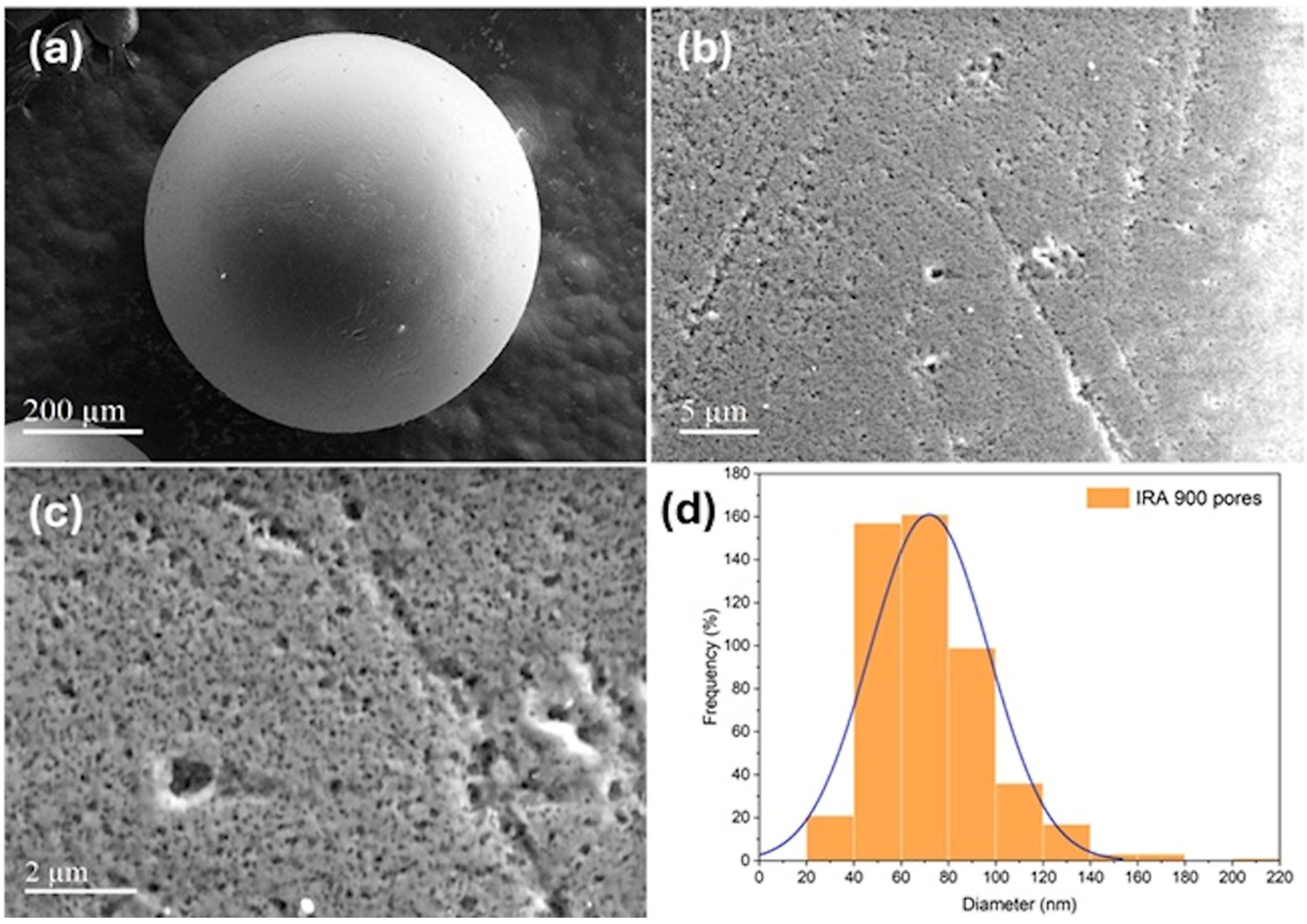

**Figure 5.** (a) SEM image of the IRA900-$HCO_3$ sample prior to the cryomilling process. (b) SEM image displaying the surface morphology of the IRA900-$HCO_3$ sample after cryomilling, where the presence of macropores is evident. (c) High-magnification SEM image of the cryomilled IRA900-$HCO_3$ sample, highlighting the variation in pore sizes across the surface. (d) Histogram of the pore size distribution obtained from SEM image analysis of cryomilled IRA900-$HCO_3$. The average pore diameter was determined to be approximately 80 nm with standard deviations of <10%.

### 3.2. Surface Enhancement with Nanowires

The morphological characteristics of the sorbents and surface-enhancing nanowires were investigated using scanning electron microscopy (SEM). This technique provides detailed insight into the physical features of both the cryomilled IRA900-$HCO_3$ and Ni-coated Ag nanowires (NWs). Figure 4a,b shows SEM micrographs of the Ag and Ni-coated Ag NWs employed as SERS substrates in Raman measurements. The optical properties of the Ni-coated Ag NWs were analyzed by using UV−vis spectroscopy. Figure 4c displays the absorbance spectra of Ag and Ni-coated Ag NWs as a function of wavelength. The blue line represents the absorbance of Ag NWs, revealing two peaks: one at 356 nm, attributed to the out-of-plane quadrupole resonance, and another at 394 nm, corresponding to the out-of-plane dipole resonance of Ag NWs. The red line illustrates the absorbance spectrum of Ni-coated Ag NWs, indicating the suppression of the transverse SPR of Ag NWs at 356 nm, which aligns with nickel deposition on the wire surfaces. A slight red shift is also observed in Ni-coated Ag NWs.[53] Figure 4d shows TEM images of the synthesized Ni-coated Ag nanowires, which display a uniform, elongated morphology with diameters ranging from approximately 80 to 120 nm and length extending up to several micrometers. Figure 4e presents a high-resolution TEM (HRTEM) image of a nanowire.

Energy-dispersive X-ray spectroscopy (EDS) analysis, shown in Figure 4f, illustrates the Ni-coated structure of the nanowires. The image is an RGB composite showing the spatial distribution of Ag (red) and Ni (green) across the nanowire. The Ag mapping is concentrated in the central region, confirming the presence of a metallic silver core. Surrounding the Ag is a distinct green ring, which signifies a coating rich in Ni. Figure 4g,h presents micrographs of the *IRA*900-$HCO_3$ cryomilled samples without and with Ni-coated Ag NWs, respectively. Finally, Figure 4i *IRA*900-$HCO_3$ shows a micrograph of the sample with Ni-coated Ag NWs, highlighting the distribution of these NWs within the sample.

Figure 5a shows a scanning electron microscopy (SEM) image of the IRA900 resin prior to cryomilling. The SEM image in Figure 5b reveals a porous surface. A higher magnification image in Figure 5c highlights the presence of macropores distributed across the surface. The pore size distribution of the IRA900 sample is presented in Figure 5d as a histogram of pore diameters (nm). A total of 498 individual pores were measured to construct the histogram, providing a statistically robust representation of the sample's morphology. The height of each bar represents the number of pores within the corresponding size range. The distribution exhibits a narrow peak with an average pore diameter of approximately 80 nm, indicative of a relatively uniform porosity. The macropore size distribution of IRA900−$HCO_3$ plays a critical role in dictating the mass transfer of $CO_2$ and $H_2O$ to and from the surface of the particle. Finer pore sizes, approaching that of the free volume fraction between the polymer strands of the resin, can only be achieved through alternative methods like $N_2$ adsorption with a pore analysis, see for example,[26,27] which detail the 1−10 nm pore domains using both nonlinear density functional theory (NLDFT) and Barrett−Joyner−Halenda BJH analyses.

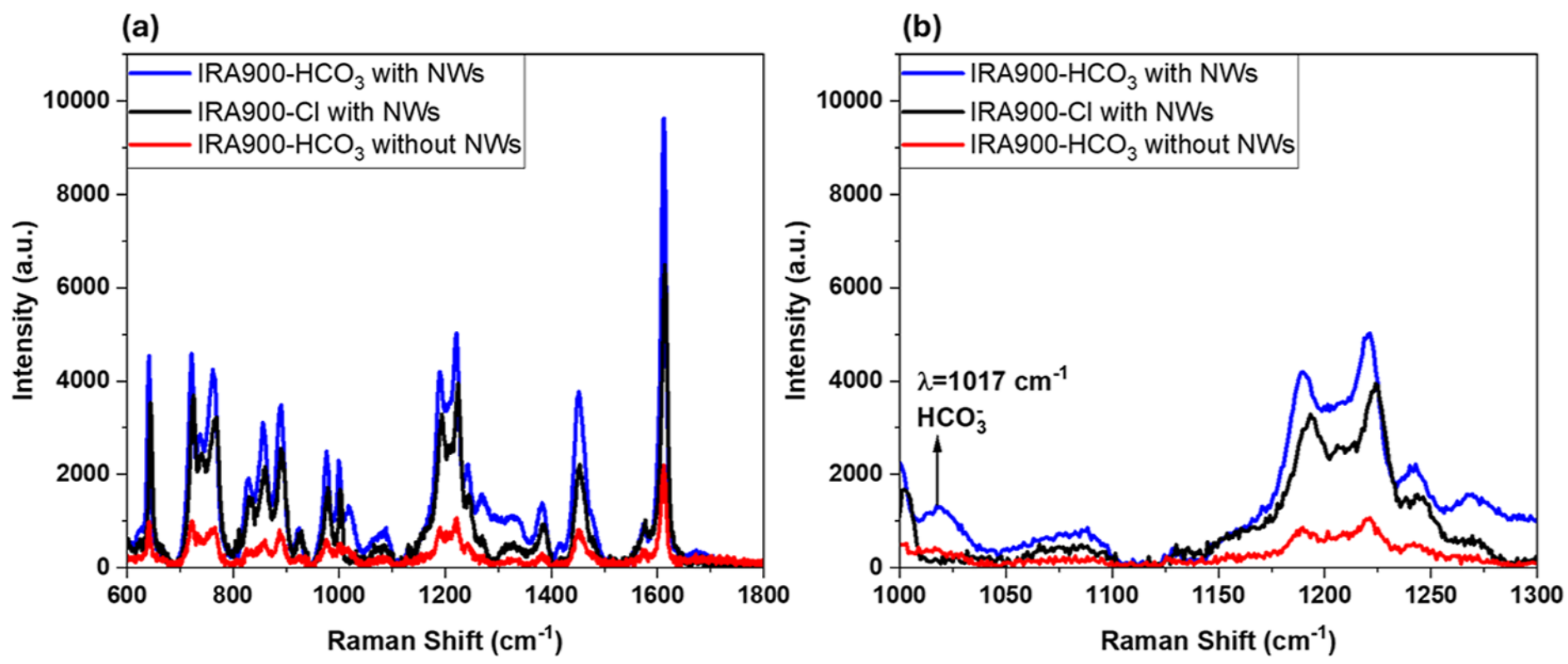

**Figure 6.** (a) Raman spectra of *IRA*900-$HCO_3$ with Ni-coated Ag NWs (blue line), the control IRA900-Cl with Ni-coated Ag NWs (black line), and *IRA*900-$HCO_3$ without Ni-coated Ag NW (red line) samples as a function of wavelength shift under dry $N_2$ conditions. (b) Spectrum of the *IRA*900-$HCO_3$ sample shows a distinct peak at 1017 $cm^{-1}$, corresponding to the characteristic band of $HCO_3^-$.

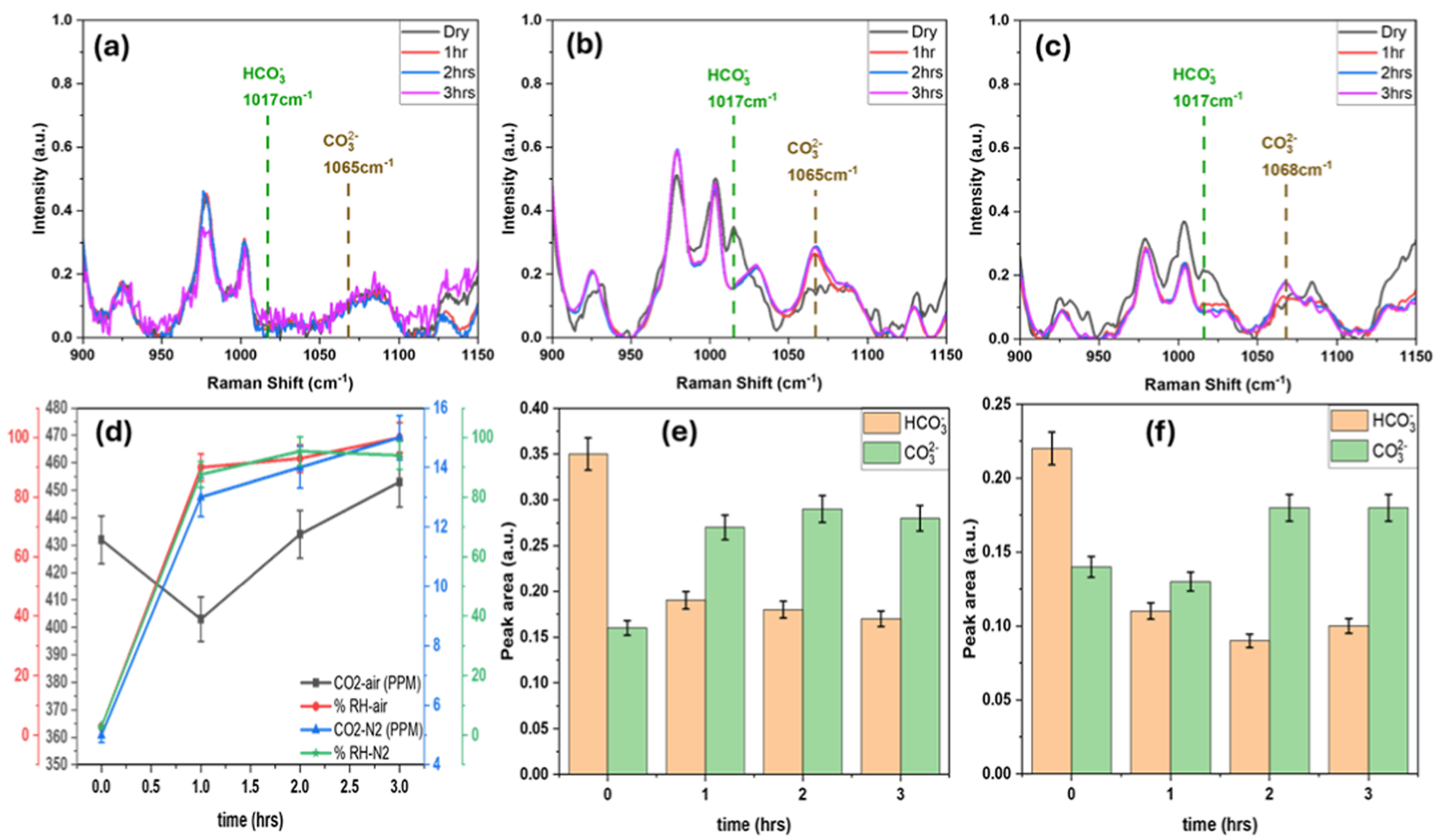

**Figure 7.** Time-evolved Raman spectra of (a) IRA900-Cl in air, (b) *IRA*900-$HCO_3$ in air, and (c) *IRA*900-$HCO_3$ in $N_2$ during a step change from low to high humidity. (d) Time series measurements of sample chamber gas composition as $CO_2$ (black) and %RH (red) in air and $CO_2$ (blue) and %RH in $N_2$ (green). (e,f) Bar graphs of integrated Raman band areas corresponding to $HCO_3^-$ and $CO_3^{2-}$ species under air (e) and $N_2$ (f). Results indicate a progressive conversion of bicarbonate to carbonate under humid conditions, particularly in an inert $N_2$ atmosphere.

The SERS effect was evaluated on both the control IRA900-Cl and the reactive IRA900-$HCO_3$ samples with and without Ni-coated Ag NWs under dry conditions. Figure 6a shows Raman measurements of *IRA*900-$HCO_3$ with (blue) and without (red) Ni-coated Ag NWs, and IRA900-Cl with Ni-coated Ag NWs (black) as a function of the Raman shift under dry $N_2$ conditions.

The Raman spectrum of the *IRA*900-$HCO_3$ samples with NWs exhibits a significant increase in Raman intensity, referred to as the enhancement factor (EF), compared to the spectrum of the *IRA*900-$HCO_3$ without. This enhancement arises from the SERS effect generated by the interaction between the Ni-coated Ag NWs, and the red laser used during the Raman measurements. The spectrum of the *IRA*900-$HCO_3$ sample shows a distinct 1017 $cm^{-1}$ peak, characteristic of $HCO_3^-$ under dry conditions, which is absent in IRA900-Cl (Figure 6b).[54,55] We note that Raman measurements in the 2800−3800 $cm^{-1}$ region, corresponding to O−H stretching modes, were not possible for IRA900−$HCO_3$ with red laser excitation (633−785 nm), as strong absorption by the polymer matrix caused localized heating and irreversible degradation. Thus, the inability to probe this region arises from the intrinsic

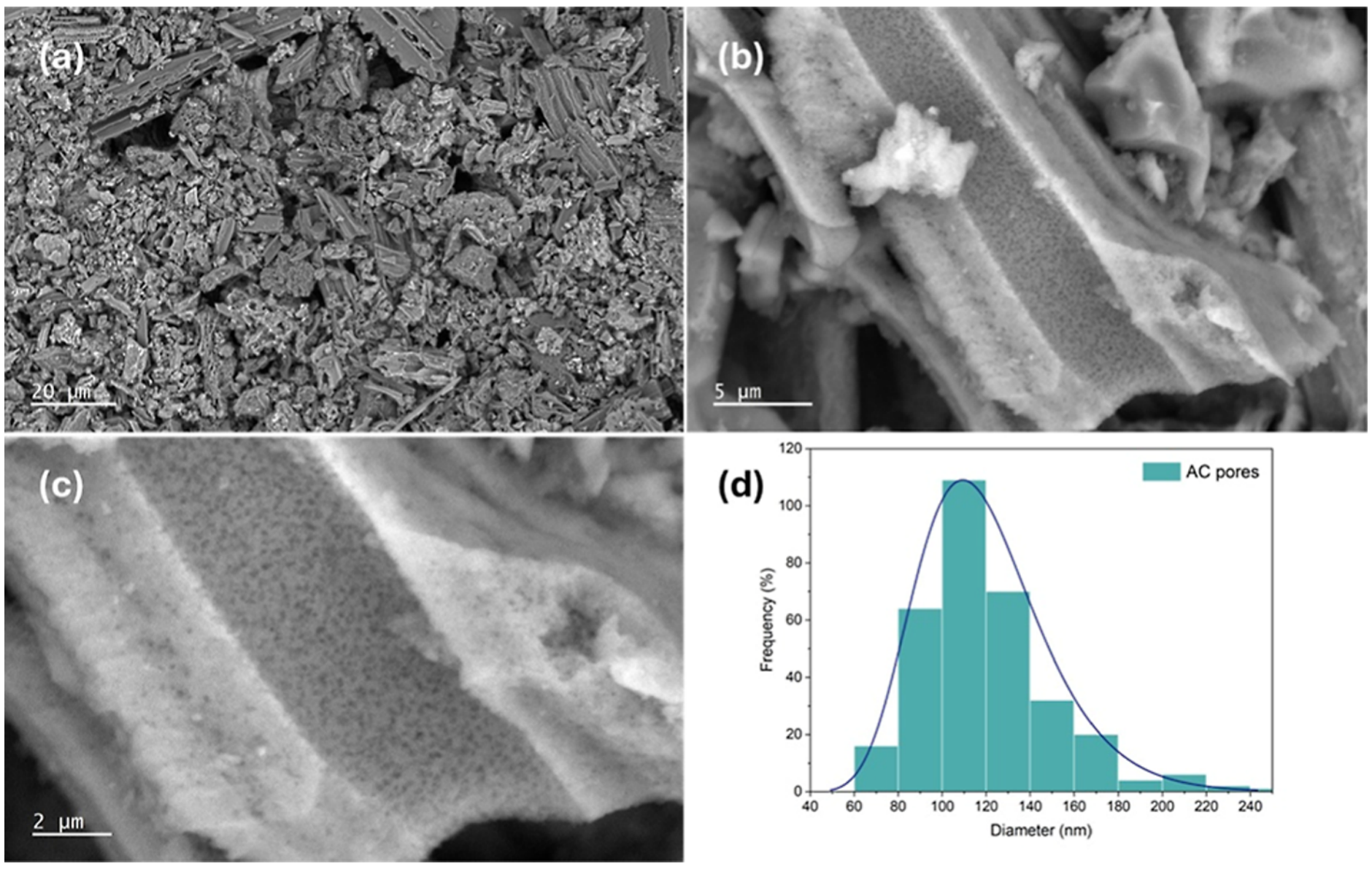

**Figure 8.** (a) Overview of the surface morphology of the activated carbon impregnated with $KHCO_3$. (b) Magnified view revealing the porous texture formed on the activated carbon walls. (c) Close-up image highlighting a uniform distribution of nanoscale pores along the walls. (d) Pore size distribution histogram derived from SEM image analysis, showing an average diameter of approximately 110 nm with standard deviations of <10%.

photothermal sensitivity of IRA900−$HCO_3$, not the absence of Raman-active modes.

To calculate the EF, eq 4 was used

$$EF = \frac{I_{SERS} \cdot n_{Raman}}{I_{Raman} \cdot n_{SERS}} \quad (4)$$

where EF is SERS enhancement factor, $I_{SERS}$ (Figure S3) is the Raman intensity using SERS, $I_{Raman}$ is the Raman intensity without using SERS, $n_{SERS}$ represents the number of molecules in contact with the SERS substrate within the probed area, while $n_{Raman}$ denotes the number of molecules present within the laser's excitation volume during standard Raman scattering measurements.[56−58] The EF was calculated based on the bicarbonate vibrational band centered at 1017 $cm^{-1}$ under dry nitrogen (Dry-$N_2$) conditions. The measured intensity of this peak was 1200 arbitrary units (au) in the presence of the SERS substrate, compared to 200 au in its absence. After performing the calculations, an EF of $3.0 \times 10^9$ was obtained. All EF calculations are detailed in eqs S2−S10.

### 3.3. $CO_2$ Reactivity in Ion-Exchange Resin

To assess anion speciation changes from $CO_2$ (de)sorption with IRA900−$HCO_3$ (eqs 1−3), Raman spectra were collected before and after a step increase in RH. Figure 7a shows control data for IRA900−Cl under dry and humid air ($PCO_2$−430 ppm) over 3 h. No time-dependent spectral changes are observed, and characteristic bicarbonate ($HCO_3^-$, 1017 $cm^{-1}$) and carbonate ($CO_3^{2-}$, 1065 $cm^{-1}$) bands are absent, confirming that the chloride system does not respond to variations in humidity.

Figure 7b,c displays the time evolution of IRA900-$HCO_3$ Raman spectra in air (b) and $N_2$ (c) while undergoing a step increase in humidity from 2.8 to 95% RH. Figure 7d shows the correlating change in gas concentration over time in the sample chamber from the initial dry state to the humidified state, with $CO_2$ reported in parts per million, ppm (black in air, blue in $N_2$) and % RH (red in air, green in $N_2$). Under dry air (420 ppm of $CO_2$), the Raman spectrum exhibits a distinct $HCO_3^-$ band at 1017 $cm^{-1}$ (35% integrated area relative to the 1615 $cm^{-1}$ peak) and a smaller $CO_3^{2-}$ band at 1065 $cm^{-1}$ (16%). Under dry $N_2$ (<10 ppm of $CO_2$), similar features are observed, though with lower relative intensities (22% for $HCO_3^-$ and 14% for $CO_3^{2-}$), yielding a lower $HCO_3^-/CO_3^{2-}$ ratio, 1.6 vs 2.2 in air. In both conditions, an increase in RH corresponds to an increase in the $CO_3^{2-}$ area and corresponding decreases in the $HCO_3^-$ area, supporting the predicted MS mechanism described by (eq 3). Error bars in Figure 7e,f correspond to the standard deviation of these measurements that have been added to reflect the spatial reproducibility of the data.

Humidification in the presence of $CO_2$ (Figure 7b,e) produces a 50% decline in the $HCO_3^-$ peak area with the fastest change within the first hour. This rate aligns with sorption kinetics observed in RH cycling experiments in 400 ppm of $CO_2$ (Figure 3). Concurrently, the intensity of the $CO_3^{2-}$ signal initially rises by 88% before declining to 75% of its starting value. Stoichiometric charge balance predicts that for every $CO_2$ molecule released, two $HCO_3^-$ are consumed per $CO_3^{2-}$ formed (eq 3), but the Raman data deviate from this expectation. One possible explanation is a hydration-induced shift of the 1017 $cm^{-1}$ band, with a new feature emerging at 1020−1050 $cm^{-1}$. This could reflect altered hydrogen bonding or an unidentified phase that convolutes the $HCO_3^-$ signal. The persistence of a residual $HCO_3^-$ band is consistent with the MS sorption data, where the capacity oscillates by only 1.1 mmol $g^{-1}$ out of 3.0 mmol $g^{-1}$ total exchange capacity, indicating incomplete participation of $HCO_3^-$ in the swing.

Hydration in $N_2$, Figure 7c,f, causes a stronger decrease in the $HCO_3^-$ peak (74%) with a smaller increase in $CO_3^{2-}$ (∼29%). This pattern better matches the stoichiometric expectation of the MS but suggests an additional process at a low $PCO_2$. Specifically, hydroxide formation may emerge under these conditions, explaining the overall (bi)carbonate signal suppression and muted carbonate response compared to air. We note again that detecting the hydroxide phase directly in the O−H stretching region (2800−3800 $cm^{-1}$) was not possible for IRA900−$HCO_3$ using red excitation (633−785 nm) due to photothermal degradation of the polymer.

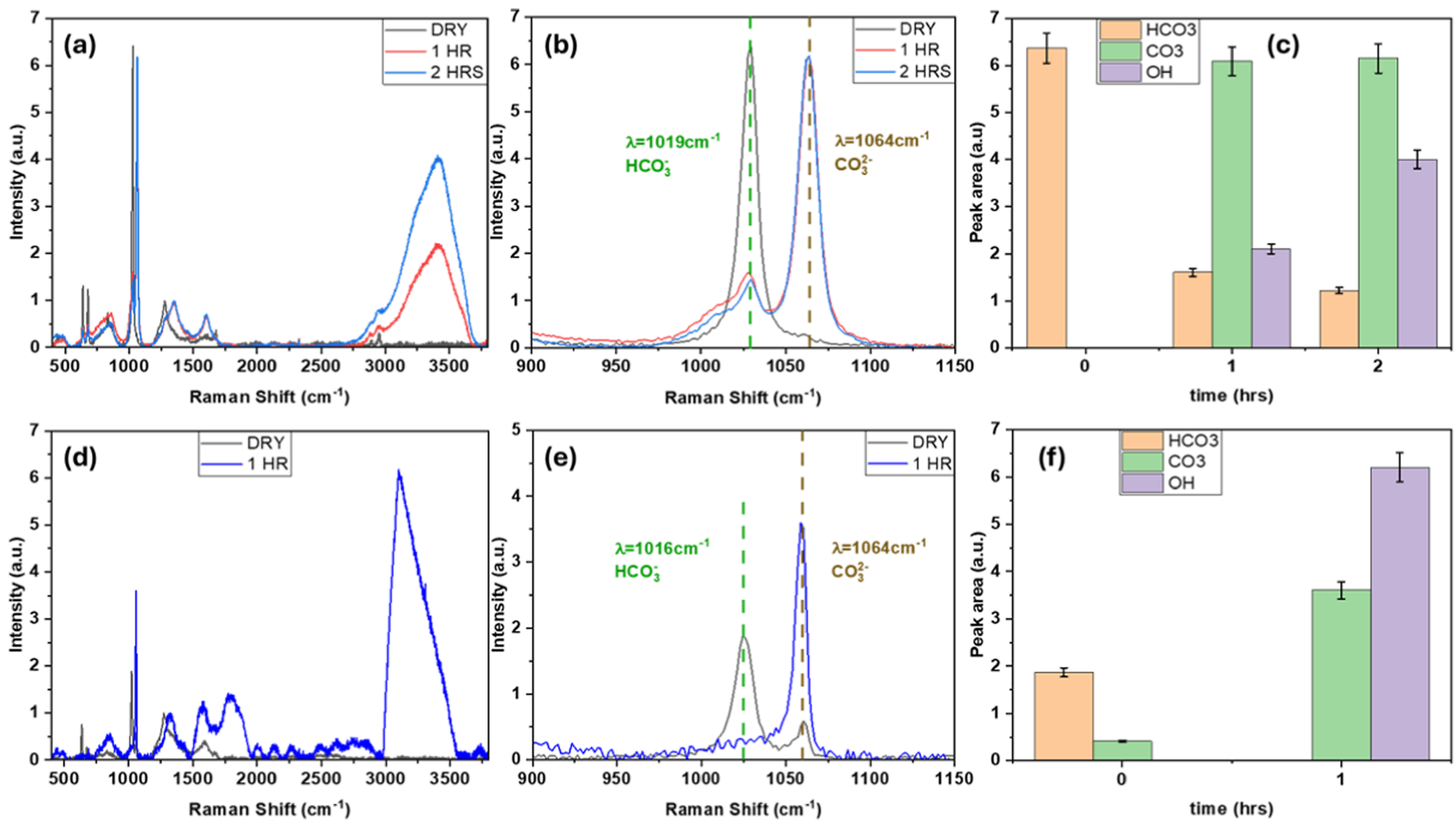

**Figure 9.** Time-dependent Raman analysis of AC−$KHCO_3$. (a−c) show spectra under humid air; (d−f) under humid $N_2$. (a) Full-range Raman spectra in dry (black) and after 1 h (red) and 2 h (blue) of humid air exposure. (b) Zoomed 900−1150 $cm^{-1}$ region showing $HCO_3^-$ decline (1019 $cm^{-1}$) and $CO_3^{2-}$ rise (1064 $cm^{-1}$). (c) Normalized peak areas of $HCO_3^-$, $CO_3^{2-}$, and $OH^-$. (d) Full-range spectra under dry (black) and 1 h humid $N_2$ (blue). (e) Zoomed region showing $HCO_3^-$ loss (1016 $cm^{-1}$) and $CO_3^{2-}$ increase (1060 $cm^{-1}$). (f) Normalized peak areas reveal rapid bicarbonate-to-carbonate conversion under humid $N_2$. Error bars represent the standard deviation from spatially averaged spectra.

### 3.4. $CO_2$ Ion Reactivity in Bicarbonate-Impregnated Porous Carbon

Like the IRA900 samples, reactive ion speciation was examined in *AC*-$KHCO_3$, while undergoing the same step increase in humidity. Figure 8a shows a morphological overview of the activated carbon material impregnated with $KHCO_3$. At higher magnification (Figure 8b), the carbon matrix displays a well-developed porous network formed by chemical activation. Further magnification (Figure 8c) reveals a homogeneous distribution of nanoscale pores, indicating uniform etching across the carbon walls.

To quantitatively characterize this morphology, a pore size distribution analysis was conducted from the SEM images (Figure 8d). A total of 420 individual pores were measured to construct the histogram, where the *y*-axis represents the count of pores within each size range. The distribution is narrowly centered around 110 nm, confirming its macroporous character. Such pores are particularly advantages for gas adsorption because they enable rapid transport of gas species to/from the sorbent surface.

Comparison with IRA-900 highlights notable textural similarities despite differences in the chemical composition. Both materials exhibit pore sizes in the macroporous regime (50−200 nm) and broad, unimodal distributions, suggesting well-connected channel networks that facilitate the mass transport of the two sorbates to the sorbent surface. These macroporous structures therefore help explain the comparable kinetics observed in the humidity-driven bicarbonate−carbonate conversion shown in Figure 3. It should be noted, however, that features below 10 nm—known to dominate the surface area of activated carbon—were not resolved here, as this lies beyond the resolution of the SEM. Details of the microporous texture of salt-impregnated activated carbon sorbents are provided in Zheng et al.[18]

Time-resolved Raman spectra of AC−$KHCO_3$ under humidified air and $N_2$ environments are shown in Figure 9. Unlike the IRA900 scans, a wider spectral window was collected, extending to 3800 $cm^{-1}$ to capture both the $HCO_3^-$/$CO_3^{2-}$ region and the O−H stretching modes. All spectra were baseline-corrected and normalized to the D band (~1350 $cm^{-1}$) to ensure consistency in comparisons. In air (Figure 9a), the O−H stretching region shows no distinct peaks in the dry state (black). After 1 h (red), the signal intensity increases to 2.1 (arbitrary units) of the normalized peak intensity and after 2 h (blue) by 4.0 (arbitrary units). Peak deconvolution in this region (Figure S4 and Table S1) identifies seven Gaussian components (2879−3622 $cm^{-1}$), which correspond to hydrogen-bonded water species, hydrogen bonded $OH^-$ (2900 $cm^{-1}$), and free $OH^-$ ions (3500 $cm^{-1}$), consistent with literature assignments.[43]

The carbonate region (900−1200 $cm^{-1}$, Figure 9b) also evolves strongly with humidity. Under dry air, only a strong $HCO_3^-$ band is observed, comprising 6.4 of the spectral intensity. Gaussian deconvolution of this region is presented in Figure S5 with peak positions and heights quantified in Table S2. After 1 h in air, this bicarbonate peak diminishes by ~75% while a strong $CO_3^{2-}$ band emerges at 1064 $cm^{-1}$ accompanied by a small $OH^-$ signal. After 2 h, bicarbonate declines slightly further, carbonate intensity plateaus, and the $OH^-$ contribution increases from ~0.4 to ~0.6 of the normalized peak area. Figure 9c summarizes these changes, showing a first sharp then slow decline in bicarbonate concurrent with carbonate formation and gradual hydroxide growth. This behavior mirrors $CO_2$ desorption in AC-$KHCO_3$ sorption experiments at 400 ppm of $CO_2$ (Figure 3), providing mechanistic evidence for the moisture-swing reactions described by eqs 1−3.

The same experiment in $N_2$ (Figure 9d,e) shows important differences. In the dry state, both $HCO_3^-$ and $CO_3^{2-}$ are already present in a ~4:1 ratio, consistent with partial bicarbonate decomposition in the absence of $CO_2$ (eq 2). After 1 h of humidified $N_2$ exposure, the bicarbonate peak disappears entirely, replaced by a much stronger $CO_3^{2-}$ signal and an emerging $OH^-$ band that is deconvoluted in Figure S6 with peak assignment in Table S3. Deconvolution of the carbonate region identifies five components (1016−1060 $cm^{-1}$; Figure S7 and Table S4), with initial relative intensities of 27.3% for $HCO_3^-$ and 6.9% for $CO_3^{2-}$. Upon hydration, the sharp growth of carbonate to ~43% and concurrent $OH^-$ formation confirm a strong driving force toward $CO_2$ desorption when $PCO_2$ concentration is very low.

A direct comparison of the two environments underscores these differences. In both cases, bicarbonate decreases while carbonate and hydroxide grow, but the transformation is faster in $N_2$, with bicarbonate fully depleted in under 1 h. The presence of carbonate even in the initial dry $N_2$ spectrum, coupled with depletion of the carbonate signal, indicates that low $PCO_2$ accelerates conversion by destabilizing bicarbonate and favoring $CO_2$ release. By contrast, atmospheric $CO_2$ (~430 ppm) in air stabilizes bicarbonate, sustaining a higher initial $HCO_3^-$ intensity and moderating the conversion rate.

Taken together, the gas sorption and in situ Raman experiments on IRA900 and AC−$KHCO_3$ provide direct evidence of the MS mechanism described in eqs 1−3, while also clarifying how hydration and $PCO_2$ govern anion speciation in distinct host materials. For IRA900, bicarbonate dominates under dry air (~420 ppm of $CO_2$) but diminishes by ~50% within the first hour of humidification, with carbonate intensity rising more strongly than stoichiometry predicts. The excess carbonate signal likely reflects the enhanced Raman scattering efficiency of $CO_3^{2-}$ rather than true ion ratios, underscoring the challenge of extracting quantitative speciation from Raman peak areas. A residual bicarbonate band persists even under extended hydration, consistent with sorption data showing incomplete cycling of exchange sites. Under $N_2$ (<10 ppm of $CO_2$), bicarbonate is less stable, declining by ~74% with only a modest carbonate increase (~29%), suggesting hydroxide formation competes at low $PCO_2$. Interpretation of hydroxide speciation, however, is constrained by the inability to probe the O−H stretching region in the polymer due to photothermal degradation.

In contrast, the AC−$KHCO_3$ system could be probed across a wider spectral window (to 3800 $cm^{-1}$), enabling simultaneous observation of carbonate, bicarbonate, and O−H stretching features. Under humid air, bicarbonate declines by ~75% after 1 h, accompanied by a strong carbonate peak and a small hydroxide signal. After 2 h, carbonate intensity plateaus while hydroxide grows, consistent with the role of $OH^-$ in eq 2. In $N_2$, the driving force for $CO_2$ desorption is more pronounced: carbonate is already present in the dry state, bicarbonate is fully depleted within 1 h, and hydroxide contributions increase. This behavior reflects Le Chatelier's principle, where the absence of gas-phase $CO_2$ destabilizes bicarbonate and accelerates conversion to carbonate and hydroxide. Unlike polymeric IRA900, the carbon matrix allows the direct detection of hydroxide vibrational modes, providing stronger mechanistic evidence that $OH^-$ participates in the hydration-dependent equilibrium. This has important implications for the expected working capacity from these sorbents in real sorption processes where $PCO_2$ can be driven down with a water vapor sweep to enhance $CO_2$ release.

Overall, both materials confirm the hydrolysis reactivity of carbonate as the core driver of the MS process, but their structural and spectroscopic properties highlight complementary aspects of the mechanism. IRA900 demonstrates how polymer-supported bicarbonate persists under ambient $CO_2$, though stoichiometric interpretation is limited by spectral bias and inaccessible O−H modes. AC−$KHCO_3$, by contrast, reveals the carbonate−bicarbonate−hydroxide interplay across the vibrational spectrum and highlights the stronger desorption-driving force under low $PCO_2$. Together, these findings validate the proposed equilibrium framework and emphasize the importance of both hydration and the $CO_2$ partial pressure in controlling sorbent performance while also pointing to methodological limitations that must be addressed in future studies.

## 4. CONCLUSION

This study demonstrates the effective use of in situ SERS to resolve ion speciation dynamics in MS sorbents under controlled humidities and gas environments. By leveraging Ni-coated Ag nanowire-based SERS substrates, we achieved substantial enhancement in vibrational signal detection, enabling real-time monitoring of carbonate, bicarbonate, and hydroxide species within anion-exchange resins (*IRA*900-$HCO_3$) and $KHCO_3^-$-impregnated activated carbon (*AC*-$KHCO_3$). The observed reversible transitions between bicarbonate and carbonate in response to water activity shifts provide direct experimental validation of the humidity-driven equilibrium central to the MS mechanism. Importantly, this work establishes SERS as a practical, nondestructive characterization tool capable of operando analysis of reactive interfaces in DAC sorbents. The methodology enables molecular-level feedback on sorbent chemistry under operating conditions, supporting the accelerated development of MS materials with improved $CO_2$ capture efficiency, regeneration kinetics, and operational stability. This platform opens avenues for quality control and material screening in scalable carbon capture technologies aligned with climate mitigation goals.

## ■ ASSOCIATED CONTENT

### Supporting Information

The Supporting Information is available free of charge at https://pubs.acs.org/doi/10.1021/acsaenm.5c00716.

> Detailed calculations of the EF for Ni-coated Ag NWs, highlighting their suitability for fabricating SERS substrates in Raman measurements; TEM and SEM analyses used to examine the structural and surface morphology of the synthesized materials; Raman spectra of the *IRA*900-$HCO_3$ and *AC*-$KHCO_3$ membranes preprocessed through baseline correction and normal-ization; and peak deconvolution using Gaussian fitting applied to resolve overlapping bands, enabling identification of specific vibrational modes and chemical interactions under different environmental conditions (PDF)

## AUTHOR INFORMATION

### Corresponding Author

**Jennifer L. Wade** − *Steve Sanghi College of Engineering, Department of Mechanical Engineering, Northern Arizona University, Flagstaff, Arizona 86011, United States;* orcid.org/0000-0001-8372-7800; Email: Jennifer.Wade@nau.edu

### Authors

**Javier Mendez Lozoya** − *Department of Applied Physics and Materials Science, Northern Arizona University, Flagstaff, Arizona 86011, United States; Instituto Tecnológico de San Luis Potosí, Tecnológico Nacional de México, San Luis Potosí 78437, Mexico*

**Estrella Solis Mata** − *Center for Materials Interfaces in Research and Applications, Northern Arizona University, Flagstaff, Arizona 86011, United States*

**J. Jesus Velazquez Salazar** − *Department of Applied Physics and Materials Science, Northern Arizona University, Flagstaff, Arizona 86011, United States; Center for Materials Interfaces in Research and Applications, Northern Arizona University, Flagstaff, Arizona 86011, United States*

**Alondra Hernandez Cedillo** − *Department of Applied Physics and Materials Science, Northern Arizona University, Flagstaff, Arizona 86011, United States; Center for Materials Interfaces in Research and Applications, Northern Arizona University, Flagstaff, Arizona 86011, United States;* orcid.org/0000-0001-7518-2644

**Miguel Jose Yacaman** − *Department of Applied Physics and Materials Science, Northern Arizona University, Flagstaff, Arizona 86011, United States; Center for Materials Interfaces in Research and Applications, Northern Arizona University, Flagstaff, Arizona 86011, United States;* orcid.org/0000-0001-5124-7116

Complete contact information is available at: https://pubs.acs.org/10.1021/acsaenm.5c00716

### Author Contributions

J.M.L. conducted the experiments, analyzed the data, and contributed to manuscript preparation. E.S.M. was responsible for the preparation of the sorbent samples evaluated in this study. J.J.V.S. carried out SEM imaging and porosity characterization of the samples. A.H.C. assisted in the synthesis of the Ni-coated Ag nanowires used in the SERS substrates for Raman analysis. M.J.Y. provided supervision and coordination of the project, manuscript editing, and imaging support. J.L.W. led the writing and project conceptualization, contributed to ion exchange processes and sorption experiments, data interpretation, and secured funding.

### Notes

The authors declare no competing financial interest.

## ACKNOWLEDGMENTS

This material is based upon work primarily supported by the U.S. Department of Energy, Office of Science, Office of Basic Energy Sciences under Award Number DE-SC0023343 (insitu Raman, Mendez, and Wade); the Army Research Office (ARO), PTE federal award no. W911NF-23-2-0014, cognitive distributed sensing in congested radio frequency environments: FREEDOM; the National Science Foundation (NSF), sponsor grant no. 2025490, the National Science Foundation (NSF) I-CORPs grant no. 2427869, NNCI: Nanotechnology Collaborative Infrastructure Southwest (NCI-SW); the Center for Materials Interfaces in Research and Applications (!MIRA!) at Northern Arizona University (NAU) for allowing them to use its facilities.

## ABBREVIATIONS

$CO_2$, carbon dioxide; MS, moisture swing; SERS, surface-enhanced Raman spectroscopy; LSPR, localized surface plasmon resonance; SPPs, surface plasmon polaritons; Ni, nickel; Ag, silver; Au, gold; Ni-coated Ag NWs, Ni-coated Ag nanowires; $N_2$, nitrogen gas; PPT, part per thousand; PPM, parts per million; MFC 1, mass flow controller 1; MFC 2, mass flow controller 2; $I_{SERS}$, Raman intensity using SERS; $I_{Raman}$, Raman intensity without using SERS; $n_{SERS}$, number of molecules adsorbed on the SERS substrate within the probed area; $n_{Raman}$, number of molecules present within the laser's excitation volume during standard Raman scattering measurements; a.u., arbitrary units; %RH, percentage of Relative Humidity; DAA, donor−acceptor−acceptor; DDAA, donor−donor−acceptor−acceptor; DA, donor−acceptor

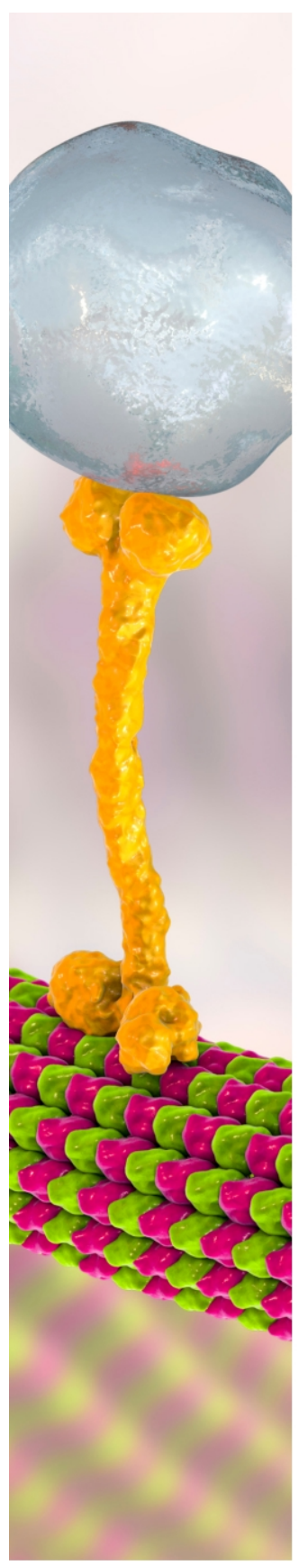